\documentclass[aps,prl,twocolumn,amsmath,superscriptaddress,amssymb,10pt]{revtex4-1}
\usepackage{graphicx}
\usepackage{ulem}
\usepackage{soul}
\usepackage{txfonts}
\usepackage[colorlinks, linkcolor=blue]{hyperref}
\usepackage{verbatim}
\usepackage{esint}
\renewcommand{\vec}[1]{\boldsymbol{#1}}

\def \r {{\vec r}}

\def \v {{\bf v}}

\def \Q{{\vec Q}}

\def \G {{\cal{G}}}

\def \beq {\begin{eqnarray}}
\def \eeq {\end{eqnarray}}
\def \tn {\textnormal}

\def \la{\langle}
\def \ra{\rangle}

\begin{document}

\title{Slow scrambling in disordered quantum systems}
\author{Brian Swingle}
\email{bswingle@stanford.edu}
\affiliation{Department of Physics, Stanford University, Palo Alto, California 94305, USA.}
\affiliation{Stanford Institute for Theoretical Physics, Stanford, California 94305, USA.}
\author{Debanjan Chowdhury}
\email{debch@mit.edu}
\affiliation{Department of Physics, Massachusetts Institute of Technology, Cambridge, Massachusetts 02139, USA.}
\begin{abstract}
Recent work has studied the growth of commutators as a probe of chaos and information scrambling in quantum many-body systems. In this work we study the effect of static disorder on the growth of commutators in a variety of contexts. We find generically that disorder slows the onset of scrambling, and, in the case of a many-body localized state, partially halts it. We access the many-body localized state using a standard fixed point Hamiltonian, and we show that operators exhibit slow logarithmic growth under time evolution. We compare the result with the expected growth of commutators in both localized and delocalized non-interacting disordered models. Finally, based on a scaling argument, we state a conjecture about the effect of weak interactions on the growth of commutators in an interacting diffusive metal.
\end{abstract}
\maketitle

{\it Introduction.-} Understanding the nature of thermalization in closed quantum systems is one of the great challenges of modern many-body physics \cite{Deutsch,Srednicki,Tasaki,rigol}, especially in light of many recent experiments probing thermalization in isolated quantum many-body systems \cite{schmied,Greiner,bloch1,bloch2,marco1,marco2,monroe1,Blatt14}. Of particular interest are the time-scales for various aspects of thermalization, from early-time relaxation to scrambling at intermediate times to the late-time build-up of complexity \cite{Susskind}. Given a many-body Hamiltonian with local interactions, relaxation describes the initial decay of local perturbations as measured by simple auto-correlation functions. Scrambling describes the slower spreading of quantum information across all the degrees of freedom of the system, rendering such information invisible to local probes. Scrambling is distinct from relaxation, with the scrambling time typically scaling in some way with the system size.

It is interesting to study the effects of static disorder on the process of thermalization because disorder is common in experimental systems and because it can lead to qualitatively new physics. In the limit of non-interacting particles, weak disorder in low dimensions or sufficiently strong disorder in three or more dimensions causes localization \cite{PWA58,Gangoffour}, which completely arrests thermalization. However, non-interacting particles, being integrable, already fail to thermalize, even when they remain delocalized. Thus it is particularly interesting to study scrambling in interacting disordered systems. It is known that the non-interacting delocalized limit can remain metallic in the presence of interactions, and recent work has shown that the localized limit is also stable to interactions \cite{BAA06,Gornyi05,prelovsek,PalHuse10,Monthus10}, resulting in many-body localization \cite{Huserev,Altmanrev}.

To probe scrambling in these systems, we study the growth of commutators of local operators. The study of such commutators is closely related to the physics of classical chaos and diagnoses a quantum version of the butterfly effect, whereby a small local perturbation eventually spreads over the entire system  \cite{ShenkerStanford2014}. To set up the precise computations, consider two local unitary operators, $V$ and $W$, along with time-evolution specified by a many-body Hamiltonian $H$. Define the squared commutator $C(t)${\footnote{For fermionic operators, it is useful to consider instead the squared anti-commutator.}} to be
\beq
C(t) = \langle [W(t),V]^\dagger [W(t),V]\rangle = 2\big(1-\tn{Re}[F(t)]\big),
\label{ct}
\eeq
where $W(t) = e^{iHt} W e^{-iHt}$ is a Heisenberg operator and where $F(t)$ is a so-called out-of-time-order (OTO) correlator
\beq
F(t) = \langle W^\dagger(t) V^\dagger W(t) V\rangle.
\eeq
Here the average $\langle ... \rangle$ is taken over any quantum state $\rho$ of interest; a natural choice, which we focus on below, is to take a product state or some short-range correlated state.

The OTO correlator $F(t)$ is our primary object of study. The physical picture is this: $W$ is meant to correspond to a simple local perturbation which grows in size and complexity under time evolution. The commutator of $W(t)$ with other simple operators $V$ diagnoses the growth of $W(t)$. The squared commutator is studied to prevent unwanted cancellations and to diagnose typical matrix elements of the commutator. We will be particularly interested in the disorder average of $F$ as well as the disorder average of $|F|^2$. We emphasize that the system (ensemble of systems) only really scrambles if, for a given disorder realization (typical disorder realization), the OTO correlator becomes small and remains small for an extended period of time.

OTO correlators first appeared many years ago in the context of semi-classical methods in superconductivity \cite{LO69}, and they have received renewed attention in the context of the AdS/CFT correspondence where they were shown to diagnose quantum chaos in black hole physics \cite{ShenkerStanford2014,MSS15,kitaevtalk}. Very recently, it has been shown how to measure OTO correlators \cite{BSexp, TGexp, BSinter} and hence scrambling. The broad relevance of scrambling in quantum many-body dynamics has also been emphasized \cite{BSexp}, for example, scrambling diagnoses the growth of quantum chaos \cite{LO69,MSS15,kitaevtalk} and the spread of entanglement \cite{hosur}.

In the spirit of investigating scrambling across a wide variety of physical systems, we study the OTO correlator $F(t)$ in simple disordered many-body models. If $W$ and $V$ are separated in space by a distance $d$, then $F(t)$ remains close to $1$ until the operator $W(t)$ has grown in space to size $d$. Our results are stated in terms of the the ``operator radius" of $W(t)$, denoted $R_W(t)$, which is defined as the distance $d$ such that $F(t)$ significantly deviates from $1$ for operators $V$ within $d$ of $W$. In a localized free-particle state, commutators simply do not grow beyond the localization length $\xi$, so $R_W(t) \sim \xi$. In a diffusive metal, commutators grow in space diffusively, with $R_W(t) \sim \sqrt{Dt}$ and $D$ the diffusion constant, but ultimately become small again at late time (with recurrences in finite size systems). Including interactions in the single-particle localized state gives a many-body localized state, and using a simple fixed point model for the many-body localized state we show that $R_W(t) \sim \xi \log(\Delta t)$ for non-conserved local operators, where $\Delta$ encodes the typical strength of interactions. This is consistent with Lieb-Robinson bounds \cite{LiebRobinson,KimLR,BurrellLR,HamzaLR}. Finally, we give a scaling argument that in a diffusive metal weak interactions lead to a ballistic growth, $R_W(t) \sim v_B t$, with a small ``butterfly velocity" \cite{BSLR,RSS15}, $v_B \sim \sqrt{D \Gamma}$ where $\Gamma$ is a small interaction induced inelastic scattering rate.

{\it Non-interacting particles.-} To orient the discussion, we first recall the behavior of commutators in non-interacting particle models. Because of Wick's theorem, all commutators of bosonic operators are ultimately proportional to the fundamental commutator (anti-commutator) of bosonic (fermionic) mode operators. We focus on fermionic operators for concreteness.

The operator $c_\r~(c_\r^\dagger)$ represents a fermionic annihilation (creation) operator at site $\r$, satisfying the usual anti-commutation algebra: $\{ c_\r, c^\dagger_{\r'}\} = \delta_{\r\r'}$. It is sufficient to study just the anti-commutator, $A_{\r\r'}(t)$, of the underlying fermion field,
\beq
A_{\r\r'}(t) = \la \{c_{\r}(t),c_{\r'}^\dagger\} \ra.
\eeq
When $A$ is small, then commutators of localized unitary operators built from the $c_{\r}$ will also be small.

As a concrete model, we consider the disordered tight-binding model \cite{PWA58} of non-interacting (NI) fermions on an infinite $d-$dimensional lattice,
\beq
H_{\tn{NI}}= -w\sum_{\langle \r \r' \rangle} \bigg(c_\r^\dagger c_{\r'} + c_{\r'}^\dagger c_\r \bigg) + \sum_\r U_\r c_\r^\dagger c_\r.
\label{ai}
\eeq
Here $w$ represents a nearest-neighbor hopping and $U_\r$ is a static random on-site potential. In one and two-dimensions, an infinitesimal disorder induces localization of the eigenstates \cite{PWA58,Gangoffour}; in three dimensions a critical amount of disorder is required to drive localization. The anti-commutator in any non-interacting fermion model is simply
\beq
A_{\r\r'}(t) = \sum_\alpha e^{-i E_\alpha t} \phi_\alpha(\r)\phi_\alpha^*(\r')
\eeq
where $\phi_\alpha(\r)$ are the single-particle energy eigenstates and $E_\alpha$ are the single-particle energies. Note that $A_{\r\r'}(t)$ is state-independent, and is simply the single-particle propagator, i.e. the amplitude for a particle to move from site $\r$ to site $\r'$ in time $t$. We will consider two cases, the localized state and the delocalized state, and leave a discussion of critical points for future work.

In the case of the localized phase, the anti-commuator never grows large. Because $A$ is a sum over single particle states, it follows that if every single-particle state is localized then $A$ remains exponentially small. Ignoring the oscillating (time-dependent) phases, which can only make $A$ smaller, we may estimate the size of $A$ by assuming that every state $\phi_\alpha$ is exponentially localized around some site $\r_\alpha$, so that the sum over $\alpha$ becomes \cite{supp}
\beq
A_{\r\r'} \lesssim \sum_\alpha e^{- |\r-\r_\alpha|/\xi - |\r'-\r_\alpha|/\xi} ~e^{i\Theta_\alpha(\r,\r')},
\eeq
where $\Theta_\alpha(\r,\r')$ represents the phases associated with the overlap of $\phi_\alpha(\r)\phi^*_\alpha(\r')$. The disorder average of $A_{\r\r'}$ (denoted $\overline{A_{\r\r'}}$) is zero as a result of averaging over the phases. However, assuming that the $\r_\alpha$ are distributed roughly uniformly in space, it immediately follows that $|\overline{A_{\r\r'}}|^2 \lesssim e^{-|\r-\r'|/\xi}$.

In the case of the delocalized phase, the resulting many-body state is a diffusive metal. Conduction of charge and heat take place, and the density-density response function exhibits a diffusion pole \cite{AS10}. Within the Born approximation, we have $\overline{A_{\r\r'}} = A_{\r\r'}^\text{clean} \exp\left(-\frac{t}{2\tau}\right)$, with the simplifying assumption of an energy-independent scattering rate $\tau$. However, as is well known, this quantity is not a good measure of the fermion motion. A better picture of the dynamics is obtained by $\overline{|A_{\r\r'}|^2}$. This calculation is closely related to the disorder average of the density-density response function \cite{AS10}, and exhibits diffusive behavior, $\overline{|A_{\r\r'}(t)|^2} \sim \exp\left(-\frac{|\r-\r'|^2}{2 D t} \right)/t^{d/2}$.

{\it Many-body localized state.-} We now consider the effects of interactions, first on the localized state. Assuming that the interacting state is a many-body localized state, we study a standard ``fixed point" Hamiltonian \cite{Altman1, HNO1, Abanin1, Imbrie16,BGSMBL} of $N$ spin-1/2s of the form
\beq
H_{\text{MBL}} &=& \sum_{\r_1} J_{\r_1} Z_{\r_1} + \frac{1}{2} \sum_{\r_1 \neq \r_2} J_{\r_1 \r_2} Z_{\r_1} Z_{\r_2} \nonumber\\
&+& \frac{1}{3!} \sum_{\r_1 \neq \r_2 \neq \r_3} J_{\r_1 \r_2 \r_3} Z_{\r_1} Z_{\r_2} Z_{\r_3}  + O(Z^4),
\eeq
where $Z_{\r}$ is the $z$ Pauli operator of spin $\r$. The couplings $J_{\r_1}$, $J_{\r_1 \r_2}$, ..., are assumed to be drawn at random from Gaussian distributions of mean zero and variance $\Delta^2_n(\r_1,...,\r_n)$ for the $n$-spin coupling $J_{\r_1, ...,\r_n}$. The variances $\Delta^2_n$ are assumed to decay with $n$ and with the separation between the $\r_i$. Such a Hamiltonian can be viewed as arising from $H_{\tn{NI}}$ by adding interactions and taking the limit $w\rightarrow 0$.

The time evolution of any local spin operator is given by precession about the $z$-axis in an effective field,
\beq
h_{\r} = J_{\r} + \sum_{\r_1 \neq \r} J_{\r \r_1} Z_{\r_1} + \frac{1}{2} \sum_{\r_2 \neq \r_1 \neq \r} J_{\r \r_1 \r_2} Z_{\r_1} Z_{\r_2} + O(Z^3),
\eeq
which is itself an operator that depends on the $z$-components of all other spins. The time evolution of any operator $O_{\r}$ on site $\r$ is given by $O_{\r}(t) = e^{i t h_{\r} Z_{\r}} O_{\r} e^{-i t h_{\r} Z_{\r}}$.
A basis of local operators is provided by the Pauli operators $\{X_{\r}, Y_{\r},Z_{\r}\}$; in terms of these operators, the $Z_{\r}$ are exactly conserved in time, while $X_{\r}$ and $Y_{\r}$ rotate into each other at a rate determined by $h_{\r}$.

To build intuition, assume first that $J_{\r_1 \r_2 \r_3}$ and all higher order terms are zero. Then a significantly non-zero commutator will develop between two operators $O_{\r_1}(t)$ and $O_{\r_2}$ after a time $t$ of order $1/J_{\r_1 \r_2}$. In a many-body localized phase, where $J_{\r_1 \r_2}$ is expected to decay exponentially with distance ($J_{\r_1 \r_2}\sim \Delta~e^{-|\r_1-\r_2|/\xi}$), it follows that a non-zero commutator will develop only after a time exponentially long in the distance between the operators. Phrased in terms of the operator radius, we have $R_W(t)\sim\xi\log(\Delta t)$, representing a slow logarithmic growth of non-conserved operators. This feature is also responsible for the logarithmic growth of entanglement in the MBL phase \cite{BPM12,Abanin1}; these are in fact related statements \cite{hosur}. Note, however, that without including higher spin interactions, the commutator is exactly periodic in time with period $2 \pi /J_{\r_1 \r_2}$, so a fixed realization of disorder only weakly scrambles.

Now consider including the neglected multi-spin interactions. It is useful to define an effective $J_{\r_1 \r_2}^{\text{eff}}= \frac{\partial h_{\r_1}}{\partial Z_{\r_2}}$ by
\beq
J_{\r_1 \r_2}^{\text{eff}}  =  J_{\r_1 \r_2} + \sum_{\r_3} J_{\r_1 \r_2 \r_3} Z_{\r_3} + \frac{1}{2} \sum_{\r_3 \r_4} J_{\r_1 \r_2 \r_3 \r_4} Z_{\r_3} Z_{\r_4} + ...,
\eeq
because this quantity is an operator that depends on the environment of other spins and can lead to dephasing.

For concreteness, consider the squared commutator $C(t)$ introduced in Eq.~\eqref{ct} above with the identification, $W\equiv X_{\r} = S_{\r}^+ + S_{\r}^-$ and $V\equiv X_{\r'} = S_{\r'}^+ + S_{\r'}^-$,
where $S^\pm_{\vec{r}} = (X_{\vec{r}} \pm i Y_{\vec{r}})/2$ are spin ladder operators. With this special choice, $W$ and $V$ are unitary and Hermitian. The time development of $S^\pm_{\r}$ is simple: $S^\pm_{\r}(t) = e^{\pm i h_{\r} t} S^\pm_{\r}$, so the OTO correlator is
\beq
F(t) &=& \left\langle \left(S^+_{\r} e^{i h_{\r} t} + S^-_{\r} e^{-i h_{\r} t}\right) X_{\r'} \left(S^+_{\r} e^{i h_{\r} t} + S^-_{\r} e^{-i h_{\r} t}\right) X_{\r'} \right\rangle, \\
F(t) &=& \left\langle S^+_{\r} S^-_{\r} e^{i J_{\r \r'}^{\text{eff}} t} + S^-_{\r} S^+_{\r} e^{- i J_{\r \r'}^{\text{eff}} t}\right\rangle.
\eeq
In the last equality we have used $X_{\r'} h_{\r}(...,Z_{\r'},...) = h_{\r}(...,-Z_{\r'},...) X_{\r'}$, $(S^\pm_{\r})^2 = 0$, and $X_{\r'}^2 = 1$ to simplify $F(t)$. Noting that $J_{\r \r'}^{\text{eff}}$ must remain inside the expectation value since it is an operator, the physics of $F(t)$ will thus be controlled by an average of a phase $e^{\pm i J^{\text{eff}}_{\r \r'} t}$ over different spin configurations. Such an average will generically lead to dephasing.

We now give some quantitative formulae for the case where $J_{\r_1 \r_2 \r_3}$ is non-zero, with mean zero and variance $\Delta^2_3(\r_1, \r_2, \r_3)$, but all higher-order multi-spin interactions are set to zero. The general case involves a trivial extension of the reported formulae and should only enhance dephasing. With only $J_{\r_1 \r_2 \r_3}$ non-zero, the effective coupling is $J_{\r \r'}^{\text{eff}} = J_{\r\r'} + \sum_{\vec{s}\neq \r, \r'} J_{\r \r' \vec{s}} Z_{\vec{s}}$.

There are two sources of randomness in $J_{\r \r'}^{\text{eff}}$, the random couplings themselves and the quantum operators $Z_{\vec{s}}$. We first analyze the effects of the quantum operators. For illustrative purposes, assume for the remainder of the calculation that the $Z_{\vec{s}}$ are uncorrelated. We believe similar results will hold for generic short-range correlated states, e.g. when studying the resulting dynamics of an initial approximately product state. We emphasize that we are not in general considering a thermal state, because thermal states are not privileged in models which fail to thermalize.

Viewing the $J$'s as fixed and again assuming the spins are in a product state, the quantum expectation value and quantum variance of $J_{\r \r'}^{\text{eff}}$ are
\beq
\langle J_{\r \r'}^{\text{eff}} \rangle &=& J_{\r \r'} + \sum_{\vec{s}} J_{\r \r' \vec{s}} \langle Z_{\vec{s}} \rangle, \\
\langle ~\left(J_{\r \r'}^{\text{eff}}\right)^2 \rangle - \langle J_{\r \r'}^{\text{eff}} \rangle^2 &=& \sum_{\vec{s}} J_{\r \r' \vec{s}}^2 (1 - \langle Z_{\vec{s}} \rangle^2).
\eeq
If we compute the quantum average of $e^{i J^{\text{eff}}_{\r \r'} t}$ by keeping only the first two cumulants, then we find
\beq
\left\langle e^{i J^{\text{eff}}_{\r \r'} t} \right\rangle = \exp\left( i t \langle J_{\r \r'}^{\text{eff}} \rangle - \frac{t^2}{2} \left\{  \sum_{\vec{s}} J_{\r \r' \vec{s}}^2 (1 - \langle Z_{\vec{s}}\rangle^2)\right\}\right).
\eeq

Let us now consider the disorder average of $F(t)$. Taking again $W= X_{\vec{r}_1}$ and $V = X_{\vec{r}_2}$, the first moment is trivial,
\beq
\overline{F(t)} = \exp\left( - \frac{1}{2} \Delta^2_2(\vec{r}_1,\vec{r}_2) ~t^2 + ... \right),
\eeq
where we have shown only the contribution from $J_{\vec{r}_1 \vec{r}_2}$ and $...$ denotes higher order terms. The second moment of $F(t)$ is more interesting, since we do not obtain trivial dephasing from disorder averaging the two-spin interaction. The key point is that the second moment involves two quantum averages over $\rho$, and hence can be thought of as a single quantum average over a two-copy system in the state $\rho \otimes \rho$. Then, analogous to the standard replica trick, the disorder average couples observables in the two copies together. The result is
\beq
\overline{|F(t)|^2} &=& 2 ~\exp\left( - 2 \Delta^2_2(\vec{r}_1,\vec{r}_2) ~t^2 \right)~\text{tr}\left\{\rho \otimes \rho ~S_{\r}^+ S_{\r}^- \otimes S_{\r}^- S_{\r}^+ \prod_{\vec{s} \neq \vec{r}_1,\vec{r}_2} G^{+}_{\vec{s}} \right\} \nonumber \\
 &+& \text{tr}\left\{\rho \otimes \rho ~S_{\r}^+ S_{\r}^- \otimes S_{\r}^+ S_{\r}^- \prod_{\vec{s} \neq \vec{r}_1,\vec{r}_2} G^{-}_{\vec{s}} \right\} \nonumber \\
 &+& \text{tr}\left\{\rho \otimes \rho ~S_{\r}^- S_{\r}^+ \otimes S_{\r}^- S_{\r}^+ \prod_{\vec{s} \neq \vec{r}_1,\vec{r}_2} G^{-}_{\vec{s}} \right\}, \\
 \text{with}~G^{\pm}_{\vec{s}} &=& \exp \left[ - \frac{\Delta^2_3(\vec{s})t^2}{2} (Z_{\vec{s}} \otimes I \pm I \otimes Z_{\vec{s}})^2 \right]
\eeq
and $\Delta_3(\vec{s}) = \Delta_3(\vec{r}_1,\vec{r}_2,\vec{s})$. If the $Z_{\vec{s}}$ are uncorrelated and if  $Z_{\vec{s}}=1$ with probability $q_{\vec{s}}$, then
\beq \label{F2moment}
\overline{|F(t)|^2} \sim \prod_{\vec{s} \neq \vec{r}_1,\vec{r}_2} \left[q_{\vec{s}}^2 + (1 - q_{\vec{s}})^2 + 2 q_{\vec{s}} (1-q_{\vec{s}}) e^{ - 2 \Delta^2_3(\vec{s})t^2 } \right]
\eeq
where we have dropped the faster decaying terms. One can obtain similar expressions for $\overline{F^2}$ and other higher moments of $F$ \cite{supp}. Provided all the disorder averaged moments of $F$ decay at late time, the commutator $C(t)$ will concentrate in probability around a late time value of $2$.

The physics of $\overline{|F(t)|^2}$ is as follows. At early times, the exponentials in Eq.~\eqref{F2moment} are close to one and $\overline{|F(t)|^2}$ is also close to one. As time passes, more and more of the exponentials decay towards zero and hence the product in Eq.~\eqref{F2moment} decays due to the multiplication of many numbers smaller than one. To say more, we must specify the form of the variance, which we take to be
\beq
\Delta_3(\vec{s}) = \Delta_3 \exp\left( -\frac{|\vec{r}_1 - \vec{r}_2|}{\xi} -\frac{|\vec{r}_1 - \vec{s}|}{\xi} - \frac{|\vec{r}_2 - \vec{s}|}{\xi} \right).
\eeq
A little geometry shows that contours of constant $u$ in the equation $|\vec{r}_1 - \vec{s}| + |\vec{r}_2 - \vec{s}| = u |\vec{r}_1 - \vec{r}_2|$ are ellipsoids, $|\vec{r}_1 - \vec{r}_2|^2 u^2 (u^2 - 1) = 4 (u^2 - 1) s_\parallel^2 + 4 u^2 s_\perp^2$, where $s_\parallel$ and $s_\perp$ denote the parallel and perpendicular components of $\vec{s}$ relative to $\vec{r}_1 - \vec{r}_2$. Note that $u \geq 1$ is required to have a solution. The $d$-dimensional volume of the ellipsoid is
\beq
\text{vol} \sim \frac{|\vec{r}_1 - \vec{r}_2|^d u (u^2-1)^{(d-1)/2}}{2^d}.
\eeq
The equation $\Delta_3(\vec{s}) t = 1$ denotes the rough boundary within which the exponentials in Eq.~\eqref{F2moment} have substantially decayed; in terms of the $u$ parameter just discussed, the solution is
\beq
u(t) = \frac{\xi}{|\vec{r}_1 - \vec{r}_2|} \log(\Delta_3 t) - 1.
\eeq
This gives a complex pattern of decay of $\overline{|F(t)|^2}$; to illustrate the basic physics, we make the simplifying assumption that $q_{\vec{s}} = 1/2$ and focus on late times, which gives
\beq
\overline{|F(t)|^2} \sim \exp\left( - a \xi^d \log^d(\Delta_3 t) \right),
\eeq
a quasi-polynomial decay in general spatial dimension $d$ and $a$ is a constant.

{\it Interacting diffusive metal.-}We now turn to the effects of interactions on the disordered but delocalized metallic state. In the non-interacting limit of a diffusive metal, commutators grow diffusively and then decay as a power law at late time. Now we sketch a simple argument that including interactions significantly modifies this behavior, leading to a commutator which grows ballistically, albeit with a small velocity in the limit of weak interactions, and which remains non-zero at late times.

We first note that ballistic growth is the generic case, and the fastest growth allowed by the Lieb-Robinson bound \cite{LiebRobinson,BSLR}. In the non-interacting limit, all operator growth is tied to the motion of particles (i.e. Wick's theorem); physically it is the statement that energy, charge, and entanglement are only carried by the single particle modes. When we include interactions, then the transport of energy, charge, and entanglement decouple and we expect more generic behavior for the motion of entanglement even if charge motion remains diffusive. Since operators must grow for entanglement to be generated, we also expect to obtain ballistic growth for generic operators. This is a physical argument, but below we sketch a simple ansatz giving ballistic growth.

Now the question is how to estimate the butterfly velocity $v_B$. First, the butterfly velocity should vanish in the limit that the interaction-induced inelastic scattering rate $\Gamma$ goes to zero. Second, the butterfly velocity must be constructed from a ratio of the relevant length- and time-scales, including the mean-free path $\ell$, the elastic scattering rate $\gamma$, and the inelastic rate $\Gamma$. Assuming no other scales are relevant, dimensional analysis gives $v_B \sim \ell \gamma f(\Gamma/\gamma)$. Further assuming that $v_B$ depends on $\ell$ and $\gamma$ only through the diffusion constant $D \sim \ell^2 \gamma$ then fixes the scaling function to be $f(x) \sim \sqrt{x}$ and gives $v_B \sim \sqrt{D \Gamma}$.

To better understand this form for $v_B$, let us imagine a pertubative calculation of the squared anti-commutator (or the commutator of some local bosonic operators) in the presence of interactions. The bare result is just $|A|^2$ which when disorder averaged gives the previously discussed diffusive form. Interactions lead to other terms involving integrals over powers of $A$ and other Green's functions. Assuming these interaction terms can be resummed \cite{DSweak} at early time to give exponential growth at roughly the inelastic rate $\Gamma$, then the interacting early time growth of the anti-commutator will be
\beq
C(\r,t) \sim \tn{exp}(\Gamma t)~ \tn{exp}(-\r^2 / 2 D t).
\eeq
Solving for $C(R_W(t),t) \sim 1$ gives $R_W^2 \sim D \Gamma t^2$ which is ballistic growth with butterfly velocity $v_B \sim \sqrt{D \Gamma}$. Based on this scaling argument, we conjecture that a full perturbative calculation will yield the same result.

{\it Discussion.-}In this work we studied the growth of operators under time evolution in disordered models using squared (anti-)commutators and OTO correlators. Two directions for future work are a systematic perturbative calculation of scrambling in the interacting diffusive metal and a study of the behavior of scrambling at the transition from a many-body localized phase to an ergodic phase \cite{VHA15,PVP15}. Another interesting direction concerns scrambling in glassy models, including long-range models \cite{SY,BSinter}, where we may study the interplay of glassy physics and scrambling.

One can also study scrambling in the Aubry-Andr$\acute{\tn{e}}$ model \cite{AA}, where in Eq.~\eqref{ai}, $U_\r=V\cos(\Q.\r)$ is an incommensurate potential with an {\it irrational} period $|\Q|/2\pi$. This model is known to have a single metal-insulator transition across the self-dual point, $w\leftrightarrow V/2$ (in one-dimension). In the presence of short-ranged interactions, the localized phase can become many-body localized \cite{interAA}, and it would be interesting to study the onset of scrambling in such a model explicitly. Precisely at the self-dual point, $V=2w$, the eigenstates of the non-interacting model are known to be `critical' and form a Cantor-set. The fate of this phase in the presence of interactions and the behavior of $F(t)$ are currently being computed \cite{BSDC}.

Our study has focused on various disorder averaged OTO correlators, but one could ask about rare-region effects \cite{TV10} as well. For example, can rare thermalized regions in the localized phase effectively give a short-cut to faster scrambling? Alternatively, in one dimension, rare localized regions in the ergodic phase should slow the growth of operators, leading to slower scrambling.

Experimentally, the effects of static disorder can be induced, for example, using laser speckle  \cite{aspect} or by modulating an optical lattice with incommensurate wavelengths, and experiments observing some of the physics of many-body localization have recently been carried out \cite{bloch1,bloch2,marco1,marco2,monroe1}. There has been a recent proposal focusing on echo-like measurements to probe the collective dephasing \cite{echo}; we have shown that OTO correlators also access the slow logarithmic growth of dephasing which is characteristic of the many-body localized state. Some of the experimental methods for adding disorder are compatibile with the time-reversal requirements of \cite{BSexp}, so measurements of scrambling might be possible. It would be particularly interesting to make measurements of OTO correlators at the transition between localized and ergodic states, where the growth of operators may diagnose the onset of ergodicity across the transition.

{\it Note added.-}During the final stages of preparation of this manuscript, three other studies \cite{xie16,zhai16,Chen16} of OTO correlators in many-body localized states appeared.

{\it Acknowledgements.-} BS is supported by the Simons Foundation and the Stanford Institute for Theoretical Physics. DC is supported by a postdoctoral fellowship from the Gordon and Betty Moore Foundation, under the EPiQS initiative, Grant GBMF-4303.

\bibliographystyle{apsrev4-1_custom}
\bibliography{scrambling}

\begin{thebibliography}{58}%
\makeatletter
\providecommand \@ifxundefined [1]{%
 \@ifx{#1\undefined}
}%
\providecommand \@ifnum [1]{%
 \ifnum #1\expandafter \@firstoftwo
 \else \expandafter \@secondoftwo
 \fi
}%
\providecommand \@ifx [1]{%
 \ifx #1\expandafter \@firstoftwo
 \else \expandafter \@secondoftwo
 \fi
}%
\providecommand \natexlab [1]{#1}%
\providecommand \enquote  [1]{``#1''}%
\providecommand \bibnamefont  [1]{#1}%
\providecommand \bibfnamefont [1]{#1}%
\providecommand \citenamefont [1]{#1}%
\providecommand \href@noop [0]{\@secondoftwo}%
\providecommand \href [0]{\begingroup \@sanitize@url \@href}%
\providecommand \@href[1]{\@@startlink{#1}\@@href}%
\providecommand \@@href[1]{\endgroup#1\@@endlink}%
\providecommand \@sanitize@url [0]{\catcode `\\12\catcode `\$12\catcode
  `\&12\catcode `\#12\catcode `\^12\catcode `\_12\catcode `\%12\relax}%
\providecommand \@@startlink[1]{}%
\providecommand \@@endlink[0]{}%
\providecommand \url  [0]{\begingroup\@sanitize@url \@url }%
\providecommand \@url [1]{\endgroup\@href {#1}{\urlprefix }}%
\providecommand \urlprefix  [0]{URL }%
\providecommand \Eprint [0]{\href }%
\providecommand \doibase [0]{http://dx.doi.org/}%
\providecommand \selectlanguage [0]{\@gobble}%
\providecommand \bibinfo  [0]{\@secondoftwo}%
\providecommand \bibfield  [0]{\@secondoftwo}%
\providecommand \translation [1]{[#1]}%
\providecommand \BibitemOpen [0]{}%
\providecommand \bibitemStop [0]{}%
\providecommand \bibitemNoStop [0]{.\EOS\space}%
\providecommand \EOS [0]{\spacefactor3000\relax}%
\providecommand \BibitemShut  [1]{\csname bibitem#1\endcsname}%
\let\auto@bib@innerbib\@empty
\bibitem [{\citenamefont {Deutsch}(1991)}]{Deutsch}%
  \BibitemOpen
  \bibfield  {author} {\bibinfo {author} {\bibfnamefont {J.~M.}\ \bibnamefont
  {Deutsch}},\ }\bibfield  {title} {\enquote {\bibinfo {title} {Quantum
  statistical mechanics in a closed system},}\ }\href {\doibase
  10.1103/PhysRevA.43.2046} {\bibfield  {journal} {\bibinfo  {journal} {Phys.
  Rev. A}\ }\textbf {\bibinfo {volume} {43}},\ \bibinfo {pages} {2046}
  (\bibinfo {year} {1991})}\BibitemShut {NoStop}%
\bibitem [{\citenamefont {Srednicki}(1994)}]{Srednicki}%
  \BibitemOpen
  \bibfield  {author} {\bibinfo {author} {\bibfnamefont {M.}~\bibnamefont
  {Srednicki}},\ }\bibfield  {title} {\enquote {\bibinfo {title} {Chaos and
  quantum thermalization},}\ }\href {\doibase 10.1103/PhysRevE.50.888}
  {\bibfield  {journal} {\bibinfo  {journal} {Phys. Rev. E}\ }\textbf {\bibinfo
  {volume} {50}},\ \bibinfo {pages} {888} (\bibinfo {year} {1994})}\BibitemShut
  {NoStop}%
\bibitem [{\citenamefont {Tasaki}(1998)}]{Tasaki}%
  \BibitemOpen
  \bibfield  {author} {\bibinfo {author} {\bibfnamefont {H.}~\bibnamefont
  {Tasaki}},\ }\bibfield  {title} {\enquote {\bibinfo {title} {From quantum
  dynamics to the canonical distribution: General picture and a rigorous
  example},}\ }\href {\doibase 10.1103/PhysRevLett.80.1373} {\bibfield
  {journal} {\bibinfo  {journal} {Phys. Rev. Lett.}\ }\textbf {\bibinfo
  {volume} {80}},\ \bibinfo {pages} {1373} (\bibinfo {year}
  {1998})}\BibitemShut {NoStop}%
\bibitem [{\citenamefont {Rigol}\ \emph {et~al.}(2008)\citenamefont {Rigol},
  \citenamefont {Dunjko},\ and\ \citenamefont {Olshanii}}]{rigol}%
  \BibitemOpen
  \bibfield  {author} {\bibinfo {author} {\bibfnamefont {M.}~\bibnamefont
  {Rigol}}, \bibinfo {author} {\bibfnamefont {V.}~\bibnamefont {Dunjko}}, \
  and\ \bibinfo {author} {\bibfnamefont {M.}~\bibnamefont {Olshanii}},\
  }\bibfield  {title} {\enquote {\bibinfo {title} {Thermalization and its
  mechanism for generic isolated quantum systems},}\ }\href
  {http://dx.doi.org/10.1038/nature06838} {\bibfield  {journal} {\bibinfo
  {journal} {Nature}\ }\textbf {\bibinfo {volume} {452}},\ \bibinfo {pages}
  {854} (\bibinfo {year} {2008})}\BibitemShut {NoStop}%
\bibitem [{\citenamefont {Langen}\ \emph {et~al.}(2013)\citenamefont {Langen},
  \citenamefont {Geiger}, \citenamefont {Kuhnert}, \citenamefont {Rauer},\ and\
  \citenamefont {Schmiedmayer}}]{schmied}%
  \BibitemOpen
  \bibfield  {author} {\bibinfo {author} {\bibfnamefont {T.}~\bibnamefont
  {Langen}}, \bibinfo {author} {\bibfnamefont {R.}~\bibnamefont {Geiger}},
  \bibinfo {author} {\bibfnamefont {M.}~\bibnamefont {Kuhnert}}, \bibinfo
  {author} {\bibfnamefont {B.}~\bibnamefont {Rauer}}, \ and\ \bibinfo {author}
  {\bibfnamefont {J.}~\bibnamefont {Schmiedmayer}},\ }\bibfield  {title}
  {\enquote {\bibinfo {title} {Local emergence of thermal correlations in an
  isolated quantum many-body system},}\ }\href
  {http://dx.doi.org/10.1038/nphys2739} {\bibfield  {journal} {\bibinfo
  {journal} {Nat Phys}\ }\textbf {\bibinfo {volume} {9}},\ \bibinfo {pages}
  {640} (\bibinfo {year} {2013})}\BibitemShut {NoStop}%
\bibitem [{\citenamefont {{Kaufman}}\ \emph {et~al.}(2016)\citenamefont
  {{Kaufman}}, \citenamefont {{Tai}}, \citenamefont {{Lukin}}, \citenamefont
  {{Rispoli}}, \citenamefont {{Schittko}}, \citenamefont {{Preiss}},\ and\
  \citenamefont {{Greiner}}}]{Greiner}%
  \BibitemOpen
  \bibfield  {author} {\bibinfo {author} {\bibfnamefont {A.~M.}\ \bibnamefont
  {{Kaufman}}}, \bibinfo {author} {\bibfnamefont {M.~E.}\ \bibnamefont
  {{Tai}}}, \bibinfo {author} {\bibfnamefont {A.}~\bibnamefont {{Lukin}}},
  \bibinfo {author} {\bibfnamefont {M.}~\bibnamefont {{Rispoli}}}, \bibinfo
  {author} {\bibfnamefont {R.}~\bibnamefont {{Schittko}}}, \bibinfo {author}
  {\bibfnamefont {P.~M.}\ \bibnamefont {{Preiss}}}, \ and\ \bibinfo {author}
  {\bibfnamefont {M.}~\bibnamefont {{Greiner}}},\ }\bibfield  {title} {\enquote
  {\bibinfo {title} {{Quantum thermalization through entanglement in an
  isolated many-body system}},}\ }\href@noop {} {\bibfield  {journal} {\bibinfo
   {journal} {ArXiv e-prints}\ } (\bibinfo {year} {2016})},\ \Eprint
  {http://arxiv.org/abs/1603.04409} {arXiv:1603.04409 [quant-ph]} \BibitemShut
  {NoStop}%
\bibitem [{\citenamefont {Schreiber}\ \emph {et~al.}(2015)\citenamefont
  {Schreiber}, \citenamefont {Hodgman}, \citenamefont {Bordia}, \citenamefont
  {L{\"u}schen}, \citenamefont {Fischer}, \citenamefont {Vosk}, \citenamefont
  {Altman}, \citenamefont {Schneider},\ and\ \citenamefont {Bloch}}]{bloch1}%
  \BibitemOpen
  \bibfield  {author} {\bibinfo {author} {\bibfnamefont {M.}~\bibnamefont
  {Schreiber}}, \bibinfo {author} {\bibfnamefont {S.~S.}\ \bibnamefont
  {Hodgman}}, \bibinfo {author} {\bibfnamefont {P.}~\bibnamefont {Bordia}},
  \bibinfo {author} {\bibfnamefont {H.~P.}\ \bibnamefont {L{\"u}schen}},
  \bibinfo {author} {\bibfnamefont {M.~H.}\ \bibnamefont {Fischer}}, \bibinfo
  {author} {\bibfnamefont {R.}~\bibnamefont {Vosk}}, \bibinfo {author}
  {\bibfnamefont {E.}~\bibnamefont {Altman}}, \bibinfo {author} {\bibfnamefont
  {U.}~\bibnamefont {Schneider}}, \ and\ \bibinfo {author} {\bibfnamefont
  {I.}~\bibnamefont {Bloch}},\ }\bibfield  {title} {\enquote {\bibinfo {title}
  {Observation of many-body localization of interacting fermions in a
  quasirandom optical lattice},}\ }\href {\doibase 10.1126/science.aaa7432}
  {\bibfield  {journal} {\bibinfo  {journal} {Science}\ }\textbf {\bibinfo
  {volume} {349}},\ \bibinfo {pages} {842} (\bibinfo {year} {2015})},\ \Eprint
  {http://arxiv.org/abs/http://science.sciencemag.org/content/349/6250/842.full.pdf}
  {http://science.sciencemag.org/content/349/6250/842.full.pdf} \BibitemShut
  {NoStop}%
\bibitem [{\citenamefont {Choi}\ \emph {et~al.}(2016)\citenamefont {Choi},
  \citenamefont {Hild}, \citenamefont {Zeiher}, \citenamefont {Schau{\ss}},
  \citenamefont {Rubio-Abadal}, \citenamefont {Yefsah}, \citenamefont
  {Khemani}, \citenamefont {Huse}, \citenamefont {Bloch},\ and\ \citenamefont
  {Gross}}]{bloch2}%
  \BibitemOpen
  \bibfield  {author} {\bibinfo {author} {\bibfnamefont {J.-y.}\ \bibnamefont
  {Choi}}, \bibinfo {author} {\bibfnamefont {S.}~\bibnamefont {Hild}}, \bibinfo
  {author} {\bibfnamefont {J.}~\bibnamefont {Zeiher}}, \bibinfo {author}
  {\bibfnamefont {P.}~\bibnamefont {Schau{\ss}}}, \bibinfo {author}
  {\bibfnamefont {A.}~\bibnamefont {Rubio-Abadal}}, \bibinfo {author}
  {\bibfnamefont {T.}~\bibnamefont {Yefsah}}, \bibinfo {author} {\bibfnamefont
  {V.}~\bibnamefont {Khemani}}, \bibinfo {author} {\bibfnamefont {D.~A.}\
  \bibnamefont {Huse}}, \bibinfo {author} {\bibfnamefont {I.}~\bibnamefont
  {Bloch}}, \ and\ \bibinfo {author} {\bibfnamefont {C.}~\bibnamefont
  {Gross}},\ }\bibfield  {title} {\enquote {\bibinfo {title} {Exploring the
  many-body localization transition in two dimensions},}\ }\href {\doibase
  10.1126/science.aaf8834} {\bibfield  {journal} {\bibinfo  {journal}
  {Science}\ }\textbf {\bibinfo {volume} {352}},\ \bibinfo {pages} {1547}
  (\bibinfo {year} {2016})},\ \Eprint
  {http://arxiv.org/abs/http://science.sciencemag.org/content/352/6293/1547.full.pdf}
  {http://science.sciencemag.org/content/352/6293/1547.full.pdf} \BibitemShut
  {NoStop}%
\bibitem [{\citenamefont {Kondov}\ \emph {et~al.}(2015)\citenamefont {Kondov},
  \citenamefont {McGehee}, \citenamefont {Xu},\ and\ \citenamefont
  {DeMarco}}]{marco1}%
  \BibitemOpen
  \bibfield  {author} {\bibinfo {author} {\bibfnamefont {S.~S.}\ \bibnamefont
  {Kondov}}, \bibinfo {author} {\bibfnamefont {W.~R.}\ \bibnamefont {McGehee}},
  \bibinfo {author} {\bibfnamefont {W.}~\bibnamefont {Xu}}, \ and\ \bibinfo
  {author} {\bibfnamefont {B.}~\bibnamefont {DeMarco}},\ }\bibfield  {title}
  {\enquote {\bibinfo {title} {Disorder-induced localization in a strongly
  correlated atomic hubbard gas},}\ }\href {\doibase
  10.1103/PhysRevLett.114.083002} {\bibfield  {journal} {\bibinfo  {journal}
  {Phys. Rev. Lett.}\ }\textbf {\bibinfo {volume} {114}},\ \bibinfo {pages}
  {083002} (\bibinfo {year} {2015})}\BibitemShut {NoStop}%
\bibitem [{\citenamefont {Meldgin}\ \emph {et~al.}(2016)\citenamefont
  {Meldgin}, \citenamefont {Ray}, \citenamefont {Russ}, \citenamefont {Chen},
  \citenamefont {Ceperley},\ and\ \citenamefont {DeMarco}}]{marco2}%
  \BibitemOpen
  \bibfield  {author} {\bibinfo {author} {\bibfnamefont {C.}~\bibnamefont
  {Meldgin}}, \bibinfo {author} {\bibfnamefont {U.}~\bibnamefont {Ray}},
  \bibinfo {author} {\bibfnamefont {P.}~\bibnamefont {Russ}}, \bibinfo {author}
  {\bibfnamefont {D.}~\bibnamefont {Chen}}, \bibinfo {author} {\bibfnamefont
  {D.~M.}\ \bibnamefont {Ceperley}}, \ and\ \bibinfo {author} {\bibfnamefont
  {B.}~\bibnamefont {DeMarco}},\ }\bibfield  {title} {\enquote {\bibinfo
  {title} {Probing the bose glass-superfluid transition using quantum quenches
  of disorder},}\ }\href {http://dx.doi.org/10.1038/nphys3695} {\bibfield
  {journal} {\bibinfo  {journal} {Nat Phys}\ }\textbf {\bibinfo {volume}
  {advance online publication}},\  (\bibinfo {year} {2016})}\BibitemShut
  {NoStop}%
\bibitem [{\citenamefont {Smith}\ \emph {et~al.}(2016)\citenamefont {Smith},
  \citenamefont {Lee}, \citenamefont {Richerme}, \citenamefont {Neyenhuis},
  \citenamefont {Hess}, \citenamefont {Hauke}, \citenamefont {Heyl},
  \citenamefont {Huse},\ and\ \citenamefont {Monroe}}]{monroe1}%
  \BibitemOpen
  \bibfield  {author} {\bibinfo {author} {\bibfnamefont {J.}~\bibnamefont
  {Smith}}, \bibinfo {author} {\bibfnamefont {A.}~\bibnamefont {Lee}}, \bibinfo
  {author} {\bibfnamefont {P.}~\bibnamefont {Richerme}}, \bibinfo {author}
  {\bibfnamefont {B.}~\bibnamefont {Neyenhuis}}, \bibinfo {author}
  {\bibfnamefont {P.~W.}\ \bibnamefont {Hess}}, \bibinfo {author}
  {\bibfnamefont {P.}~\bibnamefont {Hauke}}, \bibinfo {author} {\bibfnamefont
  {M.}~\bibnamefont {Heyl}}, \bibinfo {author} {\bibfnamefont {D.~A.}\
  \bibnamefont {Huse}}, \ and\ \bibinfo {author} {\bibfnamefont
  {C.}~\bibnamefont {Monroe}},\ }\bibfield  {title} {\enquote {\bibinfo {title}
  {Many-body localization in a quantum simulator with programmable random
  disorder},}\ }\href {http://dx.doi.org/10.1038/nphys3783} {\bibfield
  {journal} {\bibinfo  {journal} {Nat Phys}\ }\textbf {\bibinfo {volume}
  {advance online publication}},\  (\bibinfo {year} {2016})}\BibitemShut
  {NoStop}%
\bibitem [{\citenamefont {Jurcevic}\ \emph {et~al.}(2014)\citenamefont
  {Jurcevic}, \citenamefont {Lanyon}, \citenamefont {Hauke}, \citenamefont
  {Hempel}, \citenamefont {Zoller}, \citenamefont {Blatt},\ and\ \citenamefont
  {Roos}}]{Blatt14}%
  \BibitemOpen
  \bibfield  {author} {\bibinfo {author} {\bibfnamefont {P.}~\bibnamefont
  {Jurcevic}}, \bibinfo {author} {\bibfnamefont {B.~P.}\ \bibnamefont
  {Lanyon}}, \bibinfo {author} {\bibfnamefont {P.}~\bibnamefont {Hauke}},
  \bibinfo {author} {\bibfnamefont {C.}~\bibnamefont {Hempel}}, \bibinfo
  {author} {\bibfnamefont {P.}~\bibnamefont {Zoller}}, \bibinfo {author}
  {\bibfnamefont {R.}~\bibnamefont {Blatt}}, \ and\ \bibinfo {author}
  {\bibfnamefont {C.~F.}\ \bibnamefont {Roos}},\ }\bibfield  {title} {\enquote
  {\bibinfo {title} {Quasiparticle engineering and entanglement propagation in
  a quantum many-body system},}\ }\href {http://dx.doi.org/10.1038/nature13461}
  {\bibfield  {journal} {\bibinfo  {journal} {Nature}\ }\textbf {\bibinfo
  {volume} {511}},\ \bibinfo {pages} {202} (\bibinfo {year}
  {2014})}\BibitemShut {NoStop}%
\bibitem [{\citenamefont {Brown}\ \emph {et~al.}(2016)\citenamefont {Brown},
  \citenamefont {Roberts}, \citenamefont {Susskind}, \citenamefont {Swingle},\
  and\ \citenamefont {Zhao}}]{Susskind}%
  \BibitemOpen
  \bibfield  {author} {\bibinfo {author} {\bibfnamefont {A.~R.}\ \bibnamefont
  {Brown}}, \bibinfo {author} {\bibfnamefont {D.~A.}\ \bibnamefont {Roberts}},
  \bibinfo {author} {\bibfnamefont {L.}~\bibnamefont {Susskind}}, \bibinfo
  {author} {\bibfnamefont {B.}~\bibnamefont {Swingle}}, \ and\ \bibinfo
  {author} {\bibfnamefont {Y.}~\bibnamefont {Zhao}},\ }\bibfield  {title}
  {\enquote {\bibinfo {title} {Holographic complexity equals bulk action?}}\
  }\href {\doibase 10.1103/PhysRevLett.116.191301} {\bibfield  {journal}
  {\bibinfo  {journal} {Phys. Rev. Lett.}\ }\textbf {\bibinfo {volume} {116}},\
  \bibinfo {pages} {191301} (\bibinfo {year} {2016})}\BibitemShut {NoStop}%
\bibitem [{\citenamefont {Anderson}(1958)}]{PWA58}%
  \BibitemOpen
  \bibfield  {author} {\bibinfo {author} {\bibfnamefont {P.~W.}\ \bibnamefont
  {Anderson}},\ }\bibfield  {title} {\enquote {\bibinfo {title} {Absence of
  diffusion in certain random lattices},}\ }\href {\doibase
  10.1103/PhysRev.109.1492} {\bibfield  {journal} {\bibinfo  {journal} {Phys.
  Rev.}\ }\textbf {\bibinfo {volume} {109}},\ \bibinfo {pages} {1492} (\bibinfo
  {year} {1958})}\BibitemShut {NoStop}%
\bibitem [{\citenamefont {Abrahams}\ \emph {et~al.}(1979)\citenamefont
  {Abrahams}, \citenamefont {Anderson}, \citenamefont {Licciardello},\ and\
  \citenamefont {Ramakrishnan}}]{Gangoffour}%
  \BibitemOpen
  \bibfield  {author} {\bibinfo {author} {\bibfnamefont {E.}~\bibnamefont
  {Abrahams}}, \bibinfo {author} {\bibfnamefont {P.~W.}\ \bibnamefont
  {Anderson}}, \bibinfo {author} {\bibfnamefont {D.~C.}\ \bibnamefont
  {Licciardello}}, \ and\ \bibinfo {author} {\bibfnamefont {T.~V.}\
  \bibnamefont {Ramakrishnan}},\ }\bibfield  {title} {\enquote {\bibinfo
  {title} {Scaling theory of localization: Absence of quantum diffusion in two
  dimensions},}\ }\href {\doibase 10.1103/PhysRevLett.42.673} {\bibfield
  {journal} {\bibinfo  {journal} {Phys. Rev. Lett.}\ }\textbf {\bibinfo
  {volume} {42}},\ \bibinfo {pages} {673} (\bibinfo {year} {1979})}\BibitemShut
  {NoStop}%
\bibitem [{\citenamefont {Basko}\ \emph {et~al.}(2006)\citenamefont {Basko},
  \citenamefont {Aleiner},\ and\ \citenamefont {Altshuler}}]{BAA06}%
  \BibitemOpen
  \bibfield  {author} {\bibinfo {author} {\bibfnamefont {D.}~\bibnamefont
  {Basko}}, \bibinfo {author} {\bibfnamefont {I.}~\bibnamefont {Aleiner}}, \
  and\ \bibinfo {author} {\bibfnamefont {B.}~\bibnamefont {Altshuler}},\
  }\bibfield  {title} {\enquote {\bibinfo {title} {Metal-€"insulator transition
  in a weakly interacting many-electron system with localized single-particle
  states},}\ }\href {\doibase http://dx.doi.org/10.1016/j.aop.2005.11.014}
  {\bibfield  {journal} {\bibinfo  {journal} {Annals of Physics}\ }\textbf
  {\bibinfo {volume} {321}},\ \bibinfo {pages} {1126 } (\bibinfo {year}
  {2006})}\BibitemShut {NoStop}%
\bibitem [{\citenamefont {Gornyi}\ \emph {et~al.}(2005)\citenamefont {Gornyi},
  \citenamefont {Mirlin},\ and\ \citenamefont {Polyakov}}]{Gornyi05}%
  \BibitemOpen
  \bibfield  {author} {\bibinfo {author} {\bibfnamefont {I.~V.}\ \bibnamefont
  {Gornyi}}, \bibinfo {author} {\bibfnamefont {A.~D.}\ \bibnamefont {Mirlin}},
  \ and\ \bibinfo {author} {\bibfnamefont {D.~G.}\ \bibnamefont {Polyakov}},\
  }\bibfield  {title} {\enquote {\bibinfo {title} {Interacting electrons in
  disordered wires: Anderson localization and low-$t$ transport},}\ }\href
  {\doibase 10.1103/PhysRevLett.95.206603} {\bibfield  {journal} {\bibinfo
  {journal} {Phys. Rev. Lett.}\ }\textbf {\bibinfo {volume} {95}},\ \bibinfo
  {pages} {206603} (\bibinfo {year} {2005})}\BibitemShut {NoStop}%
\bibitem [{\citenamefont {\ifmmode \check{Z}\else
  \v{Z}\fi{}nidari\ifmmode~\check{c}\else \v{c}\fi{}}\ \emph
  {et~al.}(2008)\citenamefont {\ifmmode \check{Z}\else
  \v{Z}\fi{}nidari\ifmmode~\check{c}\else \v{c}\fi{}}, \citenamefont {Prosen},\
  and\ \citenamefont {Prelov\ifmmode~\check{s}\else \v{s}\fi{}ek}}]{prelovsek}%
  \BibitemOpen
  \bibfield  {author} {\bibinfo {author} {\bibfnamefont {M.}~\bibnamefont
  {\ifmmode \check{Z}\else \v{Z}\fi{}nidari\ifmmode~\check{c}\else
  \v{c}\fi{}}}, \bibinfo {author} {\bibfnamefont {T.~c.~v.}\ \bibnamefont
  {Prosen}}, \ and\ \bibinfo {author} {\bibfnamefont {P.}~\bibnamefont
  {Prelov\ifmmode~\check{s}\else \v{s}\fi{}ek}},\ }\bibfield  {title} {\enquote
  {\bibinfo {title} {Many-body localization in the heisenberg $xxz$ magnet in a
  random field},}\ }\href {\doibase 10.1103/PhysRevB.77.064426} {\bibfield
  {journal} {\bibinfo  {journal} {Phys. Rev. B}\ }\textbf {\bibinfo {volume}
  {77}},\ \bibinfo {pages} {064426} (\bibinfo {year} {2008})}\BibitemShut
  {NoStop}%
\bibitem [{\citenamefont {Pal}\ and\ \citenamefont {Huse}(2010)}]{PalHuse10}%
  \BibitemOpen
  \bibfield  {author} {\bibinfo {author} {\bibfnamefont {A.}~\bibnamefont
  {Pal}}\ and\ \bibinfo {author} {\bibfnamefont {D.~A.}\ \bibnamefont {Huse}},\
  }\bibfield  {title} {\enquote {\bibinfo {title} {Many-body localization phase
  transition},}\ }\href {\doibase 10.1103/PhysRevB.82.174411} {\bibfield
  {journal} {\bibinfo  {journal} {Phys. Rev. B}\ }\textbf {\bibinfo {volume}
  {82}},\ \bibinfo {pages} {174411} (\bibinfo {year} {2010})}\BibitemShut
  {NoStop}%
\bibitem [{\citenamefont {Monthus}\ and\ \citenamefont
  {Garel}(2010)}]{Monthus10}%
  \BibitemOpen
  \bibfield  {author} {\bibinfo {author} {\bibfnamefont {C.}~\bibnamefont
  {Monthus}}\ and\ \bibinfo {author} {\bibfnamefont {T.}~\bibnamefont
  {Garel}},\ }\bibfield  {title} {\enquote {\bibinfo {title} {Many-body
  localization transition in a lattice model of interacting fermions:
  Statistics of renormalized hoppings in configuration space},}\ }\href
  {\doibase 10.1103/PhysRevB.81.134202} {\bibfield  {journal} {\bibinfo
  {journal} {Phys. Rev. B}\ }\textbf {\bibinfo {volume} {81}},\ \bibinfo
  {pages} {134202} (\bibinfo {year} {2010})}\BibitemShut {NoStop}%
\bibitem [{\citenamefont {Nandkishore}\ and\ \citenamefont
  {Huse}(2015)}]{Huserev}%
  \BibitemOpen
  \bibfield  {author} {\bibinfo {author} {\bibfnamefont {R.}~\bibnamefont
  {Nandkishore}}\ and\ \bibinfo {author} {\bibfnamefont {D.~A.}\ \bibnamefont
  {Huse}},\ }\bibfield  {title} {\enquote {\bibinfo {title} {Many-body
  localization and thermalization in quantum statistical mechanics},}\ }\href
  {\doibase 10.1146/annurev-conmatphys-031214-014726} {\bibfield  {journal}
  {\bibinfo  {journal} {Annual Review of Condensed Matter Physics}\ }\textbf
  {\bibinfo {volume} {6}},\ \bibinfo {pages} {15} (\bibinfo {year} {2015})},\
  \Eprint
  {http://arxiv.org/abs/http://dx.doi.org/10.1146/annurev-conmatphys-031214-014726}
  {http://dx.doi.org/10.1146/annurev-conmatphys-031214-014726} \BibitemShut
  {NoStop}%
\bibitem [{\citenamefont {Altman}\ and\ \citenamefont
  {Vosk}(2015)}]{Altmanrev}%
  \BibitemOpen
  \bibfield  {author} {\bibinfo {author} {\bibfnamefont {E.}~\bibnamefont
  {Altman}}\ and\ \bibinfo {author} {\bibfnamefont {R.}~\bibnamefont {Vosk}},\
  }\bibfield  {title} {\enquote {\bibinfo {title} {Universal dynamics and
  renormalization in many-body-localized systems},}\ }\href {\doibase
  10.1146/annurev-conmatphys-031214-014701} {\bibfield  {journal} {\bibinfo
  {journal} {Annual Review of Condensed Matter Physics}\ }\textbf {\bibinfo
  {volume} {6}},\ \bibinfo {pages} {383} (\bibinfo {year} {2015})},\ \Eprint
  {http://arxiv.org/abs/http://dx.doi.org/10.1146/annurev-conmatphys-031214-014701}
  {http://dx.doi.org/10.1146/annurev-conmatphys-031214-014701} \BibitemShut
  {NoStop}%
\bibitem [{\citenamefont {Shenker}\ and\ \citenamefont
  {Stanford}(2014)}]{ShenkerStanford2014}%
  \BibitemOpen
  \bibfield  {author} {\bibinfo {author} {\bibfnamefont {S.~H.}\ \bibnamefont
  {Shenker}}\ and\ \bibinfo {author} {\bibfnamefont {D.}~\bibnamefont
  {Stanford}},\ }\bibfield  {title} {\enquote {\bibinfo {title} {Black holes
  and the butterfly effect},}\ }\href {\doibase 10.1007/JHEP03(2014)067}
  {\bibfield  {journal} {\bibinfo  {journal} {Journal of High Energy Physics}\
  }\textbf {\bibinfo {volume} {2014}},\ \bibinfo {pages} {1} (\bibinfo {year}
  {2014})}\BibitemShut {NoStop}%
\bibitem [{Note1()}]{Note1}%
  \BibitemOpen
  \bibinfo {note} {For fermionic operators, it is useful to consider instead
  the squared anti-commutator.}\BibitemShut {Stop}%
\bibitem [{\citenamefont {Larkin}\ and\ \citenamefont
  {Ovchinnikov}(1969)}]{LO69}%
  \BibitemOpen
  \bibfield  {author} {\bibinfo {author} {\bibfnamefont {A.}~\bibnamefont
  {Larkin}}\ and\ \bibinfo {author} {\bibfnamefont {Y.~N.}\ \bibnamefont
  {Ovchinnikov}},\ }\bibfield  {title} {\enquote {\bibinfo {title}
  {Quasiclassical method in the theory of superconductivity},}\ }\href@noop {}
  {\bibfield  {journal} {\bibinfo  {journal} {Soviet Journal of Experimental
  and Theoretical Physics}\ }\textbf {\bibinfo {volume} {28}},\ \bibinfo
  {pages} {1200} (\bibinfo {year} {1969})}\BibitemShut {NoStop}%
\bibitem [{\citenamefont {{Maldacena}}\ \emph {et~al.}(2015)\citenamefont
  {{Maldacena}}, \citenamefont {{Shenker}},\ and\ \citenamefont
  {{Stanford}}}]{MSS15}%
  \BibitemOpen
  \bibfield  {author} {\bibinfo {author} {\bibfnamefont {J.}~\bibnamefont
  {{Maldacena}}}, \bibinfo {author} {\bibfnamefont {S.~H.}\ \bibnamefont
  {{Shenker}}}, \ and\ \bibinfo {author} {\bibfnamefont {D.}~\bibnamefont
  {{Stanford}}},\ }\bibfield  {title} {\enquote {\bibinfo {title} {{A bound on
  chaos}},}\ }\href@noop {} {\bibfield  {journal} {\bibinfo  {journal} {ArXiv
  e-prints}\ } (\bibinfo {year} {2015})},\ \Eprint
  {http://arxiv.org/abs/1503.01409} {arXiv:1503.01409 [hep-th]} \BibitemShut
  {NoStop}%
\bibitem [{\citenamefont {Kitaev}(2014)}]{kitaevtalk}%
  \BibitemOpen
  \bibfield  {author} {\bibinfo {author} {\bibfnamefont {A.}~\bibnamefont
  {Kitaev}},\ }\bibfield  {title} {\enquote {\bibinfo {title} {Hidden
  correlations in the hawking radiation and thermal noise},}\ }in\ \href@noop
  {} {\emph {\bibinfo {booktitle} {Talk given at the Fundamental Physics Prize
  Symposium}}},\ Vol.~\bibinfo {volume} {10}\ (\bibinfo {year}
  {2014})\BibitemShut {NoStop}%
\bibitem [{\citenamefont {{Swingle}}\ \emph {et~al.}(2016)\citenamefont
  {{Swingle}}, \citenamefont {{Bentsen}}, \citenamefont {{Schleier-Smith}},\
  and\ \citenamefont {{Hayden}}}]{BSexp}%
  \BibitemOpen
  \bibfield  {author} {\bibinfo {author} {\bibfnamefont {B.}~\bibnamefont
  {{Swingle}}}, \bibinfo {author} {\bibfnamefont {G.}~\bibnamefont
  {{Bentsen}}}, \bibinfo {author} {\bibfnamefont {M.}~\bibnamefont
  {{Schleier-Smith}}}, \ and\ \bibinfo {author} {\bibfnamefont
  {P.}~\bibnamefont {{Hayden}}},\ }\bibfield  {title} {\enquote {\bibinfo
  {title} {{Measuring the scrambling of quantum information}},}\ }\href@noop {}
  {\bibfield  {journal} {\bibinfo  {journal} {ArXiv e-prints}\ } (\bibinfo
  {year} {2016})},\ \Eprint {http://arxiv.org/abs/1602.06271} {arXiv:1602.06271
  [quant-ph]} \BibitemShut {NoStop}%
\bibitem [{\citenamefont {{Zhu}}\ \emph {et~al.}(2016)\citenamefont {{Zhu}},
  \citenamefont {{Hafezi}},\ and\ \citenamefont {{Grover}}}]{TGexp}%
  \BibitemOpen
  \bibfield  {author} {\bibinfo {author} {\bibfnamefont {G.}~\bibnamefont
  {{Zhu}}}, \bibinfo {author} {\bibfnamefont {M.}~\bibnamefont {{Hafezi}}}, \
  and\ \bibinfo {author} {\bibfnamefont {T.}~\bibnamefont {{Grover}}},\
  }\bibfield  {title} {\enquote {\bibinfo {title} {{Measurement of many-body
  chaos using a quantum clock}},}\ }\href@noop {} {\bibfield  {journal}
  {\bibinfo  {journal} {ArXiv e-prints}\ } (\bibinfo {year} {2016})},\ \Eprint
  {http://arxiv.org/abs/1607.00079} {arXiv:1607.00079 [quant-ph]} \BibitemShut
  {NoStop}%
\bibitem [{\citenamefont {{Yao}}\ \emph {et~al.}(2016)\citenamefont {{Yao}},
  \citenamefont {{Grusdt}}, \citenamefont {{Swingle}}, \citenamefont {{Lukin}},
  \citenamefont {{Stamper-Kurn}}, \citenamefont {{Moore}},\ and\ \citenamefont
  {{Demler}}}]{BSinter}%
  \BibitemOpen
  \bibfield  {author} {\bibinfo {author} {\bibfnamefont {N.~Y.}\ \bibnamefont
  {{Yao}}}, \bibinfo {author} {\bibfnamefont {F.}~\bibnamefont {{Grusdt}}},
  \bibinfo {author} {\bibfnamefont {B.}~\bibnamefont {{Swingle}}}, \bibinfo
  {author} {\bibfnamefont {M.~D.}\ \bibnamefont {{Lukin}}}, \bibinfo {author}
  {\bibfnamefont {D.~M.}\ \bibnamefont {{Stamper-Kurn}}}, \bibinfo {author}
  {\bibfnamefont {J.~E.}\ \bibnamefont {{Moore}}}, \ and\ \bibinfo {author}
  {\bibfnamefont {E.~A.}\ \bibnamefont {{Demler}}},\ }\bibfield  {title}
  {\enquote {\bibinfo {title} {{Interferometric Approach to Probing Fast
  Scrambling}},}\ }\href@noop {} {\bibfield  {journal} {\bibinfo  {journal}
  {ArXiv e-prints}\ } (\bibinfo {year} {2016})},\ \Eprint
  {http://arxiv.org/abs/1607.01801} {arXiv:1607.01801 [quant-ph]} \BibitemShut
  {NoStop}%
\bibitem [{\citenamefont {Hosur}\ \emph {et~al.}(2016)\citenamefont {Hosur},
  \citenamefont {Qi}, \citenamefont {Roberts},\ and\ \citenamefont
  {Yoshida}}]{hosur}%
  \BibitemOpen
  \bibfield  {author} {\bibinfo {author} {\bibfnamefont {P.}~\bibnamefont
  {Hosur}}, \bibinfo {author} {\bibfnamefont {X.-L.}\ \bibnamefont {Qi}},
  \bibinfo {author} {\bibfnamefont {D.~A.}\ \bibnamefont {Roberts}}, \ and\
  \bibinfo {author} {\bibfnamefont {B.}~\bibnamefont {Yoshida}},\ }\bibfield
  {title} {\enquote {\bibinfo {title} {Chaos in quantum channels},}\ }\href
  {\doibase 10.1007/JHEP02(2016)004} {\bibfield  {journal} {\bibinfo  {journal}
  {Journal of High Energy Physics}\ }\textbf {\bibinfo {volume} {2016}},\
  \bibinfo {pages} {1} (\bibinfo {year} {2016})}\BibitemShut {NoStop}%
\bibitem [{\citenamefont {Lieb}\ and\ \citenamefont
  {Robinson}(1972)}]{LiebRobinson}%
  \BibitemOpen
  \bibfield  {author} {\bibinfo {author} {\bibfnamefont {E.~H.}\ \bibnamefont
  {Lieb}}\ and\ \bibinfo {author} {\bibfnamefont {D.~W.}\ \bibnamefont
  {Robinson}},\ }\bibfield  {title} {\enquote {\bibinfo {title} {The finite
  group velocity of quantum spin systems},}\ }\href {\doibase
  10.1007/BF01645779} {\bibfield  {journal} {\bibinfo  {journal}
  {Communications in Mathematical Physics}\ }\textbf {\bibinfo {volume} {28}},\
  \bibinfo {pages} {251} (\bibinfo {year} {1972})}\BibitemShut {NoStop}%
\bibitem [{\citenamefont {{Kim}}\ \emph {et~al.}(2014)\citenamefont {{Kim}},
  \citenamefont {{Chandran}},\ and\ \citenamefont {{Abanin}}}]{KimLR}%
  \BibitemOpen
  \bibfield  {author} {\bibinfo {author} {\bibfnamefont {I.~H.}\ \bibnamefont
  {{Kim}}}, \bibinfo {author} {\bibfnamefont {A.}~\bibnamefont {{Chandran}}}, \
  and\ \bibinfo {author} {\bibfnamefont {D.~A.}\ \bibnamefont {{Abanin}}},\
  }\bibfield  {title} {\enquote {\bibinfo {title} {{Local integrals of motion
  and the logarithmic lightcone in many-body localized systems}},}\ }\href@noop
  {} {\bibfield  {journal} {\bibinfo  {journal} {ArXiv e-prints}\ } (\bibinfo
  {year} {2014})},\ \Eprint {http://arxiv.org/abs/1412.3073} {arXiv:1412.3073
  [cond-mat.dis-nn]} \BibitemShut {NoStop}%
\bibitem [{\citenamefont {Burrell}\ and\ \citenamefont
  {Osborne}(2007)}]{BurrellLR}%
  \BibitemOpen
  \bibfield  {author} {\bibinfo {author} {\bibfnamefont {C.~K.}\ \bibnamefont
  {Burrell}}\ and\ \bibinfo {author} {\bibfnamefont {T.~J.}\ \bibnamefont
  {Osborne}},\ }\bibfield  {title} {\enquote {\bibinfo {title} {Bounds on the
  speed of information propagation in disordered quantum spin chains},}\ }\href
  {\doibase 10.1103/PhysRevLett.99.167201} {\bibfield  {journal} {\bibinfo
  {journal} {Phys. Rev. Lett.}\ }\textbf {\bibinfo {volume} {99}},\ \bibinfo
  {pages} {167201} (\bibinfo {year} {2007})}\BibitemShut {NoStop}%
\bibitem [{\citenamefont {Hamza}\ \emph {et~al.}(2012)\citenamefont {Hamza},
  \citenamefont {Sims},\ and\ \citenamefont {Stolz}}]{HamzaLR}%
  \BibitemOpen
  \bibfield  {author} {\bibinfo {author} {\bibfnamefont {E.}~\bibnamefont
  {Hamza}}, \bibinfo {author} {\bibfnamefont {R.}~\bibnamefont {Sims}}, \ and\
  \bibinfo {author} {\bibfnamefont {G.}~\bibnamefont {Stolz}},\ }\bibfield
  {title} {\enquote {\bibinfo {title} {Dynamical localization in disordered
  quantum spin systems},}\ }\href {\doibase 10.1007/s00220-012-1544-6}
  {\bibfield  {journal} {\bibinfo  {journal} {Communications in Mathematical
  Physics}\ }\textbf {\bibinfo {volume} {315}},\ \bibinfo {pages} {215}
  (\bibinfo {year} {2012})}\BibitemShut {NoStop}%
\bibitem [{\citenamefont {{Roberts}}\ and\ \citenamefont
  {{Swingle}}(2016)}]{BSLR}%
  \BibitemOpen
  \bibfield  {author} {\bibinfo {author} {\bibfnamefont {D.~A.}\ \bibnamefont
  {{Roberts}}}\ and\ \bibinfo {author} {\bibfnamefont {B.}~\bibnamefont
  {{Swingle}}},\ }\bibfield  {title} {\enquote {\bibinfo {title}
  {{Lieb-Robinson and the butterfly effect}},}\ }\href@noop {} {\bibfield
  {journal} {\bibinfo  {journal} {ArXiv e-prints}\ } (\bibinfo {year}
  {2016})},\ \Eprint {http://arxiv.org/abs/1603.09298} {arXiv:1603.09298
  [hep-th]} \BibitemShut {NoStop}%
\bibitem [{\citenamefont {Roberts}\ \emph {et~al.}(2015)\citenamefont
  {Roberts}, \citenamefont {Stanford},\ and\ \citenamefont {Susskind}}]{RSS15}%
  \BibitemOpen
  \bibfield  {author} {\bibinfo {author} {\bibfnamefont {D.~A.}\ \bibnamefont
  {Roberts}}, \bibinfo {author} {\bibfnamefont {D.}~\bibnamefont {Stanford}}, \
  and\ \bibinfo {author} {\bibfnamefont {L.}~\bibnamefont {Susskind}},\
  }\bibfield  {title} {\enquote {\bibinfo {title} {Localized shocks},}\
  }\href@noop {} {\bibfield  {journal} {\bibinfo  {journal} {Journal of High
  Energy Physics}\ }\textbf {\bibinfo {volume} {2015}},\ \bibinfo {pages} {1}
  (\bibinfo {year} {2015})}\BibitemShut {NoStop}%
\bibitem [{sup()}]{supp}%
  \BibitemOpen
  \href@noop {} {}\bibinfo {note} {See supplementary information for more
  details.}\BibitemShut {Stop}%
\bibitem [{\citenamefont {Altland}\ and\ \citenamefont {Simons}(2010)}]{AS10}%
  \BibitemOpen
  \bibfield  {author} {\bibinfo {author} {\bibfnamefont {A.}~\bibnamefont
  {Altland}}\ and\ \bibinfo {author} {\bibfnamefont {B.~D.}\ \bibnamefont
  {Simons}},\ }\href@noop {} {\emph {\bibinfo {title} {Condensed matter field
  theory}}}\ (\bibinfo  {publisher} {Cambridge University Press},\ \bibinfo
  {year} {2010})\BibitemShut {NoStop}%
\bibitem [{\citenamefont {Vosk}\ and\ \citenamefont {Altman}(2013)}]{Altman1}%
  \BibitemOpen
  \bibfield  {author} {\bibinfo {author} {\bibfnamefont {R.}~\bibnamefont
  {Vosk}}\ and\ \bibinfo {author} {\bibfnamefont {E.}~\bibnamefont {Altman}},\
  }\bibfield  {title} {\enquote {\bibinfo {title} {Many-body localization in
  one dimension as a dynamical renormalization group fixed point},}\ }\href
  {\doibase 10.1103/PhysRevLett.110.067204} {\bibfield  {journal} {\bibinfo
  {journal} {Phys. Rev. Lett.}\ }\textbf {\bibinfo {volume} {110}},\ \bibinfo
  {pages} {067204} (\bibinfo {year} {2013})}\BibitemShut {NoStop}%
\bibitem [{\citenamefont {Huse}\ \emph {et~al.}(2014)\citenamefont {Huse},
  \citenamefont {Nandkishore},\ and\ \citenamefont {Oganesyan}}]{HNO1}%
  \BibitemOpen
  \bibfield  {author} {\bibinfo {author} {\bibfnamefont {D.~A.}\ \bibnamefont
  {Huse}}, \bibinfo {author} {\bibfnamefont {R.}~\bibnamefont {Nandkishore}}, \
  and\ \bibinfo {author} {\bibfnamefont {V.}~\bibnamefont {Oganesyan}},\
  }\bibfield  {title} {\enquote {\bibinfo {title} {Phenomenology of fully
  many-body-localized systems},}\ }\href {\doibase 10.1103/PhysRevB.90.174202}
  {\bibfield  {journal} {\bibinfo  {journal} {Phys. Rev. B}\ }\textbf {\bibinfo
  {volume} {90}},\ \bibinfo {pages} {174202} (\bibinfo {year}
  {2014})}\BibitemShut {NoStop}%
\bibitem [{\citenamefont {Serbyn}\ \emph {et~al.}(2013)\citenamefont {Serbyn},
  \citenamefont {Papi\ifmmode~\acute{c}\else \'{c}\fi{}},\ and\ \citenamefont
  {Abanin}}]{Abanin1}%
  \BibitemOpen
  \bibfield  {author} {\bibinfo {author} {\bibfnamefont {M.}~\bibnamefont
  {Serbyn}}, \bibinfo {author} {\bibfnamefont {Z.}~\bibnamefont
  {Papi\ifmmode~\acute{c}\else \'{c}\fi{}}}, \ and\ \bibinfo {author}
  {\bibfnamefont {D.~A.}\ \bibnamefont {Abanin}},\ }\bibfield  {title}
  {\enquote {\bibinfo {title} {Local conservation laws and the structure of the
  many-body localized states},}\ }\href {\doibase
  10.1103/PhysRevLett.111.127201} {\bibfield  {journal} {\bibinfo  {journal}
  {Phys. Rev. Lett.}\ }\textbf {\bibinfo {volume} {111}},\ \bibinfo {pages}
  {127201} (\bibinfo {year} {2013})}\BibitemShut {NoStop}%
\bibitem [{\citenamefont {Imbrie}(2016)}]{Imbrie16}%
  \BibitemOpen
  \bibfield  {author} {\bibinfo {author} {\bibfnamefont {J.~Z.}\ \bibnamefont
  {Imbrie}},\ }\bibfield  {title} {\enquote {\bibinfo {title} {On many-body
  localization for quantum spin chains},}\ }\href {\doibase
  10.1007/s10955-016-1508-x} {\bibfield  {journal} {\bibinfo  {journal}
  {Journal of Statistical Physics}\ }\textbf {\bibinfo {volume} {163}},\
  \bibinfo {pages} {998} (\bibinfo {year} {2016})}\BibitemShut {NoStop}%
\bibitem [{\citenamefont {{Swingle}}(2013)}]{BGSMBL}%
  \BibitemOpen
  \bibfield  {author} {\bibinfo {author} {\bibfnamefont {B.}~\bibnamefont
  {{Swingle}}},\ }\bibfield  {title} {\enquote {\bibinfo {title} {{A simple
  model of many-body localization}},}\ }\href@noop {} {\bibfield  {journal}
  {\bibinfo  {journal} {ArXiv e-prints}\ } (\bibinfo {year} {2013})},\ \Eprint
  {http://arxiv.org/abs/1307.0507} {arXiv:1307.0507 [cond-mat.dis-nn]}
  \BibitemShut {NoStop}%
\bibitem [{\citenamefont {Bardarson}\ \emph {et~al.}(2012)\citenamefont
  {Bardarson}, \citenamefont {Pollmann},\ and\ \citenamefont {Moore}}]{BPM12}%
  \BibitemOpen
  \bibfield  {author} {\bibinfo {author} {\bibfnamefont {J.~H.}\ \bibnamefont
  {Bardarson}}, \bibinfo {author} {\bibfnamefont {F.}~\bibnamefont {Pollmann}},
  \ and\ \bibinfo {author} {\bibfnamefont {J.~E.}\ \bibnamefont {Moore}},\
  }\bibfield  {title} {\enquote {\bibinfo {title} {Unbounded growth of
  entanglement in models of many-body localization},}\ }\href {\doibase
  10.1103/PhysRevLett.109.017202} {\bibfield  {journal} {\bibinfo  {journal}
  {Phys. Rev. Lett.}\ }\textbf {\bibinfo {volume} {109}},\ \bibinfo {pages}
  {017202} (\bibinfo {year} {2012})}\BibitemShut {NoStop}%
\bibitem [{\citenamefont {{Stanford}}(2015)}]{DSweak}%
  \BibitemOpen
  \bibfield  {author} {\bibinfo {author} {\bibfnamefont {D.}~\bibnamefont
  {{Stanford}}},\ }\bibfield  {title} {\enquote {\bibinfo {title} {{Many-body
  chaos at weak coupling}},}\ }\href@noop {} {\bibfield  {journal} {\bibinfo
  {journal} {ArXiv e-prints}\ } (\bibinfo {year} {2015})},\ \Eprint
  {http://arxiv.org/abs/1512.07687} {arXiv:1512.07687 [hep-th]} \BibitemShut
  {NoStop}%
\bibitem [{\citenamefont {Vosk}\ \emph {et~al.}(2015)\citenamefont {Vosk},
  \citenamefont {Huse},\ and\ \citenamefont {Altman}}]{VHA15}%
  \BibitemOpen
  \bibfield  {author} {\bibinfo {author} {\bibfnamefont {R.}~\bibnamefont
  {Vosk}}, \bibinfo {author} {\bibfnamefont {D.~A.}\ \bibnamefont {Huse}}, \
  and\ \bibinfo {author} {\bibfnamefont {E.}~\bibnamefont {Altman}},\
  }\bibfield  {title} {\enquote {\bibinfo {title} {Theory of the many-body
  localization transition in one-dimensional systems},}\ }\href {\doibase
  10.1103/PhysRevX.5.031032} {\bibfield  {journal} {\bibinfo  {journal} {Phys.
  Rev. X}\ }\textbf {\bibinfo {volume} {5}},\ \bibinfo {pages} {031032}
  (\bibinfo {year} {2015})}\BibitemShut {NoStop}%
\bibitem [{\citenamefont {Potter}\ \emph {et~al.}(2015)\citenamefont {Potter},
  \citenamefont {Vasseur},\ and\ \citenamefont {Parameswaran}}]{PVP15}%
  \BibitemOpen
  \bibfield  {author} {\bibinfo {author} {\bibfnamefont {A.~C.}\ \bibnamefont
  {Potter}}, \bibinfo {author} {\bibfnamefont {R.}~\bibnamefont {Vasseur}}, \
  and\ \bibinfo {author} {\bibfnamefont {S.~A.}\ \bibnamefont {Parameswaran}},\
  }\bibfield  {title} {\enquote {\bibinfo {title} {Universal properties of
  many-body delocalization transitions},}\ }\href {\doibase
  10.1103/PhysRevX.5.031033} {\bibfield  {journal} {\bibinfo  {journal} {Phys.
  Rev. X}\ }\textbf {\bibinfo {volume} {5}},\ \bibinfo {pages} {031033}
  (\bibinfo {year} {2015})}\BibitemShut {NoStop}%
\bibitem [{\citenamefont {Sachdev}\ and\ \citenamefont {Ye}(1993)}]{SY}%
  \BibitemOpen
  \bibfield  {author} {\bibinfo {author} {\bibfnamefont {S.}~\bibnamefont
  {Sachdev}}\ and\ \bibinfo {author} {\bibfnamefont {J.}~\bibnamefont {Ye}},\
  }\bibfield  {title} {\enquote {\bibinfo {title} {Gapless spin-fluid ground
  state in a random quantum heisenberg magnet},}\ }\href {\doibase
  10.1103/PhysRevLett.70.3339} {\bibfield  {journal} {\bibinfo  {journal}
  {Phys. Rev. Lett.}\ }\textbf {\bibinfo {volume} {70}},\ \bibinfo {pages}
  {3339} (\bibinfo {year} {1993})}\BibitemShut {NoStop}%
\bibitem [{\citenamefont {Aubry}\ and\ \citenamefont {Andr{\'e}}(1980)}]{AA}%
  \BibitemOpen
  \bibfield  {author} {\bibinfo {author} {\bibfnamefont {S.}~\bibnamefont
  {Aubry}}\ and\ \bibinfo {author} {\bibfnamefont {G.}~\bibnamefont
  {Andr{\'e}}},\ }\bibfield  {title} {\enquote {\bibinfo {title} {Analyticity
  breaking and anderson localization in incommensurate lattices},}\ }\href@noop
  {} {\bibfield  {journal} {\bibinfo  {journal} {Ann. Israel Phys. Soc}\
  }\textbf {\bibinfo {volume} {3}},\ \bibinfo {pages} {18} (\bibinfo {year}
  {1980})}\BibitemShut {NoStop}%
\bibitem [{\citenamefont {Mastropietro}(2015)}]{interAA}%
  \BibitemOpen
  \bibfield  {author} {\bibinfo {author} {\bibfnamefont {V.}~\bibnamefont
  {Mastropietro}},\ }\bibfield  {title} {\enquote {\bibinfo {title}
  {Localization of interacting fermions in the aubry-andr\'e model},}\ }\href
  {\doibase 10.1103/PhysRevLett.115.180401} {\bibfield  {journal} {\bibinfo
  {journal} {Phys. Rev. Lett.}\ }\textbf {\bibinfo {volume} {115}},\ \bibinfo
  {pages} {180401} (\bibinfo {year} {2015})}\BibitemShut {NoStop}%
\bibitem [{\citenamefont {Swingle}\ and\ \citenamefont
  {Chowdhury}(2016)}]{BSDC}%
  \BibitemOpen
  \bibfield  {author} {\bibinfo {author} {\bibfnamefont {B.}~\bibnamefont
  {Swingle}}\ and\ \bibinfo {author} {\bibfnamefont {D.}~\bibnamefont
  {Chowdhury}},\ }\href@noop {} {} (\bibinfo {year} {2016}),\ \bibinfo {note}
  {unpublished}\BibitemShut {NoStop}%
\bibitem [{\citenamefont {Vojta}(2010)}]{TV10}%
  \BibitemOpen
  \bibfield  {author} {\bibinfo {author} {\bibfnamefont {T.}~\bibnamefont
  {Vojta}},\ }\bibfield  {title} {\enquote {\bibinfo {title} {Quantum griffiths
  effects and smeared phase transitions in metals: Theory and experiment},}\
  }\href {\doibase 10.1007/s10909-010-0205-4} {\bibfield  {journal} {\bibinfo
  {journal} {Journal of Low Temperature Physics}\ }\textbf {\bibinfo {volume}
  {161}},\ \bibinfo {pages} {299} (\bibinfo {year} {2010})}\BibitemShut
  {NoStop}%
\bibitem [{\citenamefont {Billy}\ \emph {et~al.}(2008)\citenamefont {Billy},
  \citenamefont {Josse}, \citenamefont {Zuo}, \citenamefont {Bernard},
  \citenamefont {Hambrecht}, \citenamefont {Lugan}, \citenamefont {Clement},
  \citenamefont {Sanchez-Palencia}, \citenamefont {Bouyer},\ and\ \citenamefont
  {Aspect}}]{aspect}%
  \BibitemOpen
  \bibfield  {author} {\bibinfo {author} {\bibfnamefont {J.}~\bibnamefont
  {Billy}}, \bibinfo {author} {\bibfnamefont {V.}~\bibnamefont {Josse}},
  \bibinfo {author} {\bibfnamefont {Z.}~\bibnamefont {Zuo}}, \bibinfo {author}
  {\bibfnamefont {A.}~\bibnamefont {Bernard}}, \bibinfo {author} {\bibfnamefont
  {B.}~\bibnamefont {Hambrecht}}, \bibinfo {author} {\bibfnamefont
  {P.}~\bibnamefont {Lugan}}, \bibinfo {author} {\bibfnamefont
  {D.}~\bibnamefont {Clement}}, \bibinfo {author} {\bibfnamefont
  {L.}~\bibnamefont {Sanchez-Palencia}}, \bibinfo {author} {\bibfnamefont
  {P.}~\bibnamefont {Bouyer}}, \ and\ \bibinfo {author} {\bibfnamefont
  {A.}~\bibnamefont {Aspect}},\ }\bibfield  {title} {\enquote {\bibinfo {title}
  {Direct observation of anderson localization of matter waves in a controlled
  disorder},}\ }\href {http://dx.doi.org/10.1038/nature07000} {\bibfield
  {journal} {\bibinfo  {journal} {Nature}\ }\textbf {\bibinfo {volume} {453}},\
  \bibinfo {pages} {891} (\bibinfo {year} {2008})}\BibitemShut {NoStop}%
\bibitem [{\citenamefont {Serbyn}\ \emph {et~al.}(2014)\citenamefont {Serbyn},
  \citenamefont {Knap}, \citenamefont {Gopalakrishnan}, \citenamefont
  {Papi\ifmmode~\acute{c}\else \'{c}\fi{}}, \citenamefont {Yao}, \citenamefont
  {Laumann}, \citenamefont {Abanin}, \citenamefont {Lukin},\ and\ \citenamefont
  {Demler}}]{echo}%
  \BibitemOpen
  \bibfield  {author} {\bibinfo {author} {\bibfnamefont {M.}~\bibnamefont
  {Serbyn}}, \bibinfo {author} {\bibfnamefont {M.}~\bibnamefont {Knap}},
  \bibinfo {author} {\bibfnamefont {S.}~\bibnamefont {Gopalakrishnan}},
  \bibinfo {author} {\bibfnamefont {Z.}~\bibnamefont
  {Papi\ifmmode~\acute{c}\else \'{c}\fi{}}}, \bibinfo {author} {\bibfnamefont
  {N.~Y.}\ \bibnamefont {Yao}}, \bibinfo {author} {\bibfnamefont {C.~R.}\
  \bibnamefont {Laumann}}, \bibinfo {author} {\bibfnamefont {D.~A.}\
  \bibnamefont {Abanin}}, \bibinfo {author} {\bibfnamefont {M.~D.}\
  \bibnamefont {Lukin}}, \ and\ \bibinfo {author} {\bibfnamefont {E.~A.}\
  \bibnamefont {Demler}},\ }\bibfield  {title} {\enquote {\bibinfo {title}
  {Interferometric probes of many-body localization},}\ }\href {\doibase
  10.1103/PhysRevLett.113.147204} {\bibfield  {journal} {\bibinfo  {journal}
  {Phys. Rev. Lett.}\ }\textbf {\bibinfo {volume} {113}},\ \bibinfo {pages}
  {147204} (\bibinfo {year} {2014})}\BibitemShut {NoStop}%
\bibitem [{\citenamefont {{Huang}}\ \emph {et~al.}(2016)\citenamefont
  {{Huang}}, \citenamefont {{Zhang}},\ and\ \citenamefont {{Chen}}}]{xie16}%
  \BibitemOpen
  \bibfield  {author} {\bibinfo {author} {\bibfnamefont {Y.}~\bibnamefont
  {{Huang}}}, \bibinfo {author} {\bibfnamefont {Y.-L.}\ \bibnamefont
  {{Zhang}}}, \ and\ \bibinfo {author} {\bibfnamefont {X.}~\bibnamefont
  {{Chen}}},\ }\bibfield  {title} {\enquote {\bibinfo {title}
  {{Out-of-Time-Ordered Correlator in Many-Body Localized Systems}},}\
  }\href@noop {} {\bibfield  {journal} {\bibinfo  {journal} {ArXiv e-prints}\ }
  (\bibinfo {year} {2016})},\ \Eprint {http://arxiv.org/abs/1608.01091}
  {arXiv:1608.01091 [cond-mat.dis-nn]} \BibitemShut {NoStop}%
\bibitem [{\citenamefont {{Fan}}\ \emph {et~al.}(2016)\citenamefont {{Fan}},
  \citenamefont {{Zhang}}, \citenamefont {{Shen}},\ and\ \citenamefont
  {{Zhai}}}]{zhai16}%
  \BibitemOpen
  \bibfield  {author} {\bibinfo {author} {\bibfnamefont {R.}~\bibnamefont
  {{Fan}}}, \bibinfo {author} {\bibfnamefont {P.}~\bibnamefont {{Zhang}}},
  \bibinfo {author} {\bibfnamefont {H.}~\bibnamefont {{Shen}}}, \ and\ \bibinfo
  {author} {\bibfnamefont {H.}~\bibnamefont {{Zhai}}},\ }\bibfield  {title}
  {\enquote {\bibinfo {title} {{Out-of-Time-Order Correlation for Many-Body
  Localization}},}\ }\href@noop {} {\bibfield  {journal} {\bibinfo  {journal}
  {ArXiv e-prints}\ } (\bibinfo {year} {2016})},\ \Eprint
  {http://arxiv.org/abs/1608.01914} {arXiv:1608.01914 [cond-mat.quant-gas]}
  \BibitemShut {NoStop}%
\bibitem [{\citenamefont {{Chen}}(2016)}]{Chen16}%
  \BibitemOpen
  \bibfield  {author} {\bibinfo {author} {\bibfnamefont {Y.}~\bibnamefont
  {{Chen}}},\ }\bibfield  {title} {\enquote {\bibinfo {title} {{Quantum
  Logarithmic Butterfly in Many Body Localization}},}\ }\href@noop {}
  {\bibfield  {journal} {\bibinfo  {journal} {ArXiv e-prints}\ } (\bibinfo
  {year} {2016})},\ \Eprint {http://arxiv.org/abs/1608.02765} {arXiv:1608.02765
  [cond-mat.dis-nn]} \BibitemShut {NoStop}%
\end{thebibliography}%
\section{Supplementary information}
\subsection{Non-interacting Anderson insulator}
Let us return to the anti-commutator, $A_{\r\r'}(t)$. It is more useful to study the Fourier transform, $A_{\r\r'}(\epsilon)$, of the above quantity,
\beq
A_{\r\r'}(\epsilon) =  \sum_{\alpha} \delta(\epsilon-E_{\alpha}) \phi_{\alpha}(\r)~ \phi_\alpha^*(\r').
\eeq
Let us now consider averaging the above quantity over all realizations of the disorder, i.e. over the different potentials $U_\r$. Then, $\overline{A_{\r\r'}(\epsilon)} = 0$, since the phase of the overlap of the wavefunctions $ \phi_{\alpha}(\r)~ \phi_\alpha^*(\r')$ depends on the potential and averages to zero. Instead, if we consider the modulus squared
\beq
|A_{\r\r'}(\epsilon)|^2 = \sum_{\alpha} \delta(\epsilon-E_{\alpha}) |\phi_{\alpha}(\r)|^2 |\phi^*_{\alpha}(\r')|^2 > 0.
\eeq
If we now consider averaging the above quantity over all realizations of disorder, $\overline{|A_{\r\r'}(\epsilon)|^2}$ will be translationally invariant and have a form
\beq
\overline{|A_{\r\r'}(\epsilon)|^2} = \G_\epsilon(|\r-\r'|).
\eeq
In the limit $|\r-\r'|\rightarrow\infty$, $\G_\epsilon(|\r-\r'|)\rightarrow \tn{exp}(-|\r-\r'|/\xi(\epsilon))$ for localized wavefunctions. In one and two spatial dimensions, $\xi(\epsilon)$ is always finite. In three spatial dimensions, $\xi(\epsilon)$ is finite (above a critical disorder strength) but upon approaching the mobility-edge, $\epsilon_c$, has a divergence of the form $\xi(\epsilon)\sim |\epsilon-\epsilon_c|^{-\nu}$.

\subsection{Higher moments in the many-body localized state}

If we wish to disorder average the following quantum expectation value,
\beq
I(k_+,k_-) = \left\langle e^{i J_{\r\r'}^{\text{eff}} t} \right\rangle^{k_+} \left\langle e^{-i J_{\r\r'}^{\text{eff}} t} \right\rangle^{k_-},
\eeq
then we must introduce $k_+ + k_-$ copies of the system in the state $\rho^{\otimes (k_+ + k_-)}$. Focusing again on the case of non-zero $J_{\r\r'\vec{s}}$, the resulting disorder average reads
\beq
\overline{I(k_+,k_-)} &=& \exp\left( -  (k_+ - k_-)^2 \frac{\Delta^2_2(\vec{r}_1,\vec{r}_2) ~t^2}{2} \right) \nonumber \\
&\times& \text{tr}\left\{ \rho^{\otimes (k_+ + k_-)} \prod_{\vec{s} \neq \r,\r'} G^{k_+,k_-}_{\vec{s}} \right\}
\eeq
where
\beq
 G^{k_+,k_-}_{\vec{s}} = \exp \left[ - \frac{\Delta^2_3(\vec{s})t^2}{2} \left( \sum_{a=1}^{k_+} Z^{(a)}_{\vec{s}} - \sum_{a=k_+ + 1}^{k_+ + k_-} Z^{(a)}_{\vec{s}}\right)^2 \right]
\eeq
and
\beq
Z^{(a)}_{\vec{s}} = I_1 \otimes ... \otimes Z_{\vec{s},a} \otimes ... \otimes I_{k_+ + k_-}.
\eeq
Similar formulae hold for multi-spin interactions involving more than three spins (assuming all the couplings in the Hamiltonian are Gaussian distributed and independent).

\end{document}